\documentclass[preprint,pra,lettersizepaper,doublespace,,superscriptaddress]{revtex4-1}
\usepackage{layout,float}
\usepackage{amsmath,amssymb}
\usepackage{graphicx}
\usepackage{dcolumn}
\usepackage{bm}
\usepackage[utf8]{inputenc} 
\usepackage{graphicx}
\usepackage{natbib}

\DeclareGraphicsExtensions{.bmp}

\newcommand{\be}{\begin{equation}}
\newcommand{\ee}{\end{equation}}

\newcommand{\bit}{\begin{itemize}}
\newcommand{\eit}{\end{itemize}}

\graphicspath{{Figures//}}

\begin{document}

\title{Thermal effects on the resonance fluorescence of doubly dressed artificial atoms}

\author{David F. Macias-Pinilla}

\affiliation{Grupo de F\'isica Te\'orica y Computacional, Escuela de F\'isica, Universidad Pedag\'ogica y Tecnol\'ogica de Colombia (UPTC), Tunja 150003, Boyac\'a, Colombia.}

\affiliation{Departament de Qu\'imica F\'isica i Anal\'itica, Universitat Jaume I, Castell\'o  12080, Spain}

\author{Hanz Y. Ram\'irez}

\email{hanz.ramirez@uptc.edu.co}

\affiliation{Grupo de F\'isica Te\'orica y Computacional, Escuela de F\'isica, Universidad Pedag\'ogica y Tecnol\'ogica de Colombia (UPTC), Tunja 150003, Boyac\'a, Colombia.}

\affiliation{Department of Materials Science and Engineering, University of Delaware, Newark 19716, Delaware, USA.}

\date{\today}

\begin{abstract}
	In this work, robustness of controlled density of optical states in doubly driven artificial atoms is studied under phonon dissipation. By using both perturbative and polaron approaches, we investigate the influence of carrier-phonon interactions on the emission properties of a two-level solid-state emitter, simultaneously coupled to two intense distinguishable lasers. Phonon decoherence effects on the emission spectra are found modest up to neon boiling temperatures ($\sim 30$ K), as compared with photon generation at the Fourier transform limit obtained in absence of lattice vibrations (zero temperature). These results show that optical switching and photonic modulation by means of double dressing, do not require ultra low temperatures for implementation, thus boosting its potential technological applications.   
\end{abstract}
\maketitle

\section{Introduction}
Quantum dots (QDs), often also denominated ``artificial atoms'', exhibit completely discretized energy states which allow selective probing of particular exciton transitions at the single photon level. Consequently, their optical properties have been a topical trend in the last two decades given their importance for fundamental research and technological applications \cite{burstein2012,review2019}.

Recent progresses in growth and manipulation of nanometric heterostructures have indeed fetched the analogy between atoms and solid-state zero dimensional systems to impressive levels, so that  in despite of emerging in the quantum optics realm decades after natural atoms, QD-based emitters for many purposes not just match, but potentially may surpass their atomic counterpart \cite{2020}.     

Particularly, their behavior as actual two-level systems when they are irradiated by a monochromatic field under resonant excitation, has been elucidated through pristine observation of the so-called Mollow triplet (between other related phenomena); a distinctive stimulated emission of fully quantized systems \cite{PhysRev.188.1969,vamivakas2009spin,Flagg2009,mollowdot3,mollowdot4,YMHeNatureNano2013}.

In the nineties, interesting schemes for enriched emission from natural atoms undergoing double excitation were proposed \cite{bichromaticstrong,doubledressbichromatic,Freedhoff,doublesubharmonics,suppresionatoms}, and resonance fluorescence spectra in agreement with theoretical predictions were soon reported \cite{doubledressexp,doubledressexp2}. However, because of experimental difficulties for doubly driving atoms or molecules related to implication of three or more states \cite{failedsuppresion,limitations,failedsuppressioncomment}, a couple of relevant phenomena remained unobserved for almost 20 years: peak suppression by pure quantum interference (no-population trapping involved), and peak-into-band broadening for monochromatic double dressing \cite{Freedhoff,suppresionatoms}. 

Additionally, along this century the capability to study strong light-matter coupling by embedding semiconductor quantum dots in cavities has impressively developed  \cite{CYLu2016,MITemitters}. Hence, QDs became an adequate testbed for pursuing those elusive quantum optical phenomena \cite{Ramirez2013,Armenios}, which were finally measured a few years ago in InGaAs/GaAs artificial atoms \cite{He2015}.    

Such an experiment was carried out at Helium cryogenic temperature (4.2 K), so that phonon effects could be mostly inhibited. Nevertheless, thinking of device applications for those effects, how feasible would it be to reproduce them at higher temperature is a crucial question.   

Decoherence associated to vibrations of the surrounding crystal lattice is unavoidable in semiconductor nanostructures. Consequently, the usefulness of QDs for realization and usage of quantum optical features in scalable technology is limited. For instance, phonon-induced dephasing is responsible for damping of exciton Rabi oscillations in artificial atoms \cite{PhysRevB.69.193302}. It has been also found that sidebands in QD resonance fluorescence spectra, exhibit a systematic broadening with increasing temperature due to interactions between the dressed excitons with the phonon reservoir \cite{Wei}.  

So far, most of the available studies on double dressed solid-state emitters have omitted this essential decoherence channel by assuming photon generation in the Fourier limit \cite{Ramirez2013,Armenios,He2015,PRBexp,resonantcomb,doublydrivenQW}, and the very few which consider phonons, still assume the non-dissipative low temperature limit and focus on phenomena like squeezing of the phonon reservoir or coupling to neighboring nanomechanical resonators, instead of addressing the temperature dependence of the photon emission \cite{squeezedbichromatic2016,bichromaticnanoresonator2017}. 

In this work, we focus on the monochromatic double dressing of a solid-state two-level system and simulate the thermal effects on its emission spectrum assuming that the emitter is embedded in a phonon reservoir, which is effectively the case for artificial atoms given their inherent many-body nature \cite{VagovPRB2011,Nahri2017}. The results presented here particularly consider a single QD, although the used model could be also applied to other type of photon sources, like localized defects or artificial molecules \cite{k-Rama-2017,Oscar2018,tin-vacancy,Nelson,dipolaritones,QDM2020}.  

This paper is structured in three main parts: in the first one, the model used to study doubly driven QDs is introduced while the second part is devoted to the exciton-phonon interaction. Emission spectra for different temperatures and approximation levels are presented in the third section. Finally, a brief summary and conclusions close the work.

\begin{figure}[h]
	\centering
	\includegraphics[width=10.5cm]{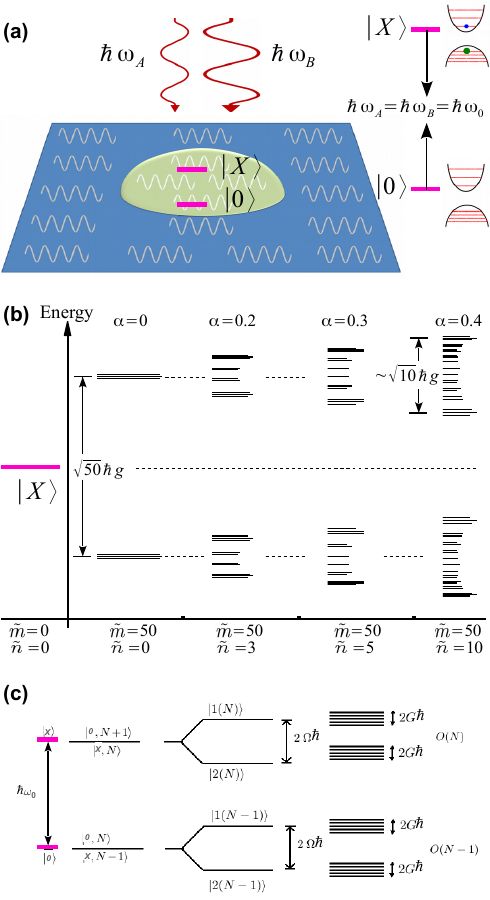}
	\caption{
		a) Schematics of the QD under resonant influence of two monochromatic lasers, b) Computed density of optical states of the energy branch corresponding to the QD state $\mid X \rangle$, evolving with the number of photons in the laser B (the line lengths represent the corresponding density magnitudes), c) Complete state configuration for the $N-1$ and $N$ rungs of the Jaynes-Cummings ladder, evolving from the bare two levels to the double dressed states.}
	\label{schematics}
\end{figure}

\section{Doubly dressed two level system}

We focus in two states of an undoped semiconductor QD: the ground state in which there is no exciton $| 0 \rangle$, and the excited state in which a localized neutral exciton is formed when an electron is promoted to the conduction band $| X \rangle$, as depicted in figure \ref{schematics}(a). Within the Jaynes-Cummings framework, the Hamiltonian for such a two-level solid-state quantum emitter with characteristic frequency $\omega_0$, simultaneously driven by two incoherent lasers of well defined frequencies $\omega_A$ and $\omega_B$, and subdue to phonon dissipation, reads

\begin{eqnarray}
\label{eq-1}
\hat{H} &=& \hat{H}_{L-M} + \hat{H}_{X-P} \hspace*{1ex} , \nonumber \\
\hat{H}_{L-M} &=& \frac{1}{2}\hbar \omega_0 \hat{\sigma}_z + \hbar \omega_A (\hat{m}+\frac{1}{2}) + \hbar g (\hat{a} \hat{\sigma}_+ + h.c.) \nonumber  + \hbar \omega_B (\hat{n}+\frac{1}{2}) + \hbar g (\hat{b} \hat{\sigma}_+ + h.c.) \hspace*{1ex} , \nonumber \\
\hat{H}_{X-P} &=& \hat{\sigma}_+ \hat{\sigma}_- \sum_{\bm{k}}  \hbar \eta_{\bm{k}} (\hat{d}_{\bm{k}} + \hat{d}_{\bm{k}}^\dagger) 
+\sum_{\bm{k}}  \hbar \omega_{\bm{k}} \hat{l}_{\bm{k}} \hspace*{1ex} ,
\label{hamiltonian}
\end{eqnarray}

where $\hat{\sigma}_+ = |X\rangle \langle 0|$ ($\hat{\sigma}_- = |0\rangle \langle X| $) is the bottom-up (top-down) dipole transition operator. In the $\hat{H}_{L-M}$ part, $g$ is the light-matter coupling, and $\hat{m}$ and $\hat{a}$ ($\hat{n}$ and $\hat{b}$) are the number of photons and photon annihilation operators for the laser A (B), respectively \cite{Ramirez2013,dipolaritones,jaynescummings,dipole-asymmetry,double-1,double-2}. Regarding the $\hat{H}_{X-P}$ part, $\eta_{\bm{\kappa}}$ is the exciton-phonon coupling for the phonon mode of frequency $\omega_{\bm{k}}$, and $\hat{l}_{\bm{k}}$ and $\hat{d}_{\bm{k}}$ are correspondingly the number of phonons and phonon annihilation operators  \cite{Oscar2018,PRL-Rabi-2010-1,PRL-Rabi-2010-2}.  

For the sake of clarity, in this section we assume a vanishing  $\eta_{\bm{k}}$. Then, the effects of a non-negligible exciton-phonon interaction are addressed in the next section.

Regarding the light-matter part of the system, an initial basis constituted by bare states of the form $| \psi_{} \rangle = | i , \tilde{m} , \tilde{n} \rangle $, where $i=0,X$ , $\tilde{m} \equiv \langle \hat{m} \rangle $ and $ \tilde{n} \equiv \langle \hat{n} \rangle$, is considered. In this paper, we study the case for resonant monochromatic double dressing, in which $\omega_{0} = \omega_{A} = \omega_{B}$. The ratio between the ``number of photons" in each laser is defined as $\alpha^2 \equiv \frac{\tilde{n}}{\tilde{m}}$ .

Then, for a given total number of photons $N \equiv \tilde{m} + \tilde{n} $, the corresponding rung of the Jaynes-Cummings ladder (JCL) is composed of $2N + 1$ states, $N + 1$ corresponding to $i=0$ and $N$ to $i=X$ (i.e. this rung is the $N$-plet \{ $ \mid 0,0,N \rangle , \mid X,0,N-1 \rangle , \mid 0,1,N-1 \rangle , \mid X,1,N-2 \rangle ,  \mid 0,2,N-2 \rangle , ... \mid X,N-2,1 \rangle , \mid 0,N-1,1 \rangle , \mid X,N-1,0 \rangle , \mid 0,N, 0 \rangle $ \} )  \cite{Ramirez2013}. 

After the first dressing, the $\left(2N+1\right)$-plet generates two groups of $N$ states (two $N$-plets), associated to the symmetric and antisymmetric combinations of the pairs of bare states coupled by the first laser, $\frac{1}{\sqrt{2}} (\mid 0 , N - \tilde{n} , \tilde{n} \rangle \pm \mid X , N - \tilde{n} - 1 , \tilde{n} \rangle) \equiv \mid \phi_{\tilde{n},\pm}^N \rangle $. Each of these combinations are the eigenstates of the 2x2 blocks in the matrix representing the emitter dressed by the laser A, whose off diagonal terms are given by 

\small
\begin{eqnarray}
\langle X , N - \tilde{n} - 1 , \tilde{n} \mid \hat{a} \hat{\sigma}_+ \mid 0 , N - \tilde{n} , \tilde{n} \rangle &\equiv& \langle X , \tilde{m} - 1 , \tilde{n} \mid \hat{a} \hat{\sigma}_+ \mid 0 , \tilde{m} , \tilde{n} \rangle \nonumber \\
&=& \langle 0 , \tilde{m} , \tilde{n} \mid \hat{a}^{\dag} \hat{\sigma}_- \mid X , \tilde{m} - 1 , \tilde{n} \rangle \hspace*{1ex} = \hbar g \sqrt{N-\tilde{n}}  \hspace*{1ex} \nonumber . \\
\label{couplinglaserA}
\end{eqnarray}  


In those $N$-plets, singly dressed states are coupled by the second laser according to


\begin{subequations}
	
	\begin{eqnarray}
	\langle\phi_{\tilde{n}}^N \pm \mid \hat{H}_{D-B} \mid \phi_{\tilde{n}'}^N \pm \rangle &=& \pm \frac{\hbar g}{2} (\sqrt{\tilde{n}'} \delta_{\tilde{n},\tilde{n}'-1} + \sqrt{\tilde{n}'+1} \delta_{\tilde{n},\tilde{n}'+1} ) \hspace*{1ex} , 
	\label{couplinglaserB1}
	\end{eqnarray}
	\begin{eqnarray}
	\langle\phi_{\tilde{n}}^N \pm \mid \hat{H}_{D-B} \mid \phi_{\tilde{n}'}^N \mp \rangle &=& \pm \frac{\hbar g}{2} (\sqrt{\tilde{n}'} \delta_{\tilde{n},\tilde{n}'-1} - \sqrt{\tilde{n}'+1} \delta_{\tilde{n},\tilde{n}'+1} ) \hspace*{1ex}  \hspace*{1ex} ,
	\label{couplinglaserB2}
	\end{eqnarray}
\end{subequations}


where the second line refers to inter $N$-plet coupling, while the first one accounts for coupling within each $N$-plet.  

The generated matrix can be numerically diagonalized to obtain the energy eigenvalues ($E_l^N ; l=1,2,...2N+1 $) and eigenstates ($\mid \psi_l^N \rangle = \sum\limits_{\tilde{n}}^{N} \sum\limits_{j=+,-} c_{\tilde{n},j}^{N,l} \mid \phi_{\tilde{n},j}^N \rangle $) of the phonon-free part of the Hamiltonian \cite{Ramirez2013}.   

Once the unitary transformation that allows change between the bare and doubly dressed bases is found, the density of optical states in terms of $\tilde{n}$ and $N$, can be calculated by using

\begin{equation}
\label{eq-5}
\rho_{\tilde{n}}^N (E)=\sum_l \sum\limits_{j=+,-} | \langle \phi_{\tilde{n},j}^{N} | \psi_l^{N} \rangle |^2 \delta(E - E_{l}^N) \hspace*{1ex}.
\end{equation}
\label{densityofstates}

Figure \ref{schematics}(b) shows the density of optical states of a JCL rung, for a fixed number of photons in the field A and increasing number of photons in the field B. There, the enrichment for possible transitions toward and from neighboring rungs is clearly appreciated. Figure \ref{schematics}(c) depicts the complete level structure for two contiguous rungs of the JCL, under double dressing.    

For the sake of physical insight it is convenient to further pursue an analytical approach. That is possible by imposing two conditions: i) the laser A being much more intense than laser B ($\tilde{m}^{2} \gg \tilde{n}^2 \Rightarrow \alpha^4 \ll 1 )$, and ii) the total number of photons being large enough so that the energy spacing between eigenvalues $E_l^N$ within a $N$-plet, is much smaller than the two involved Rabi frequencies ($N \rightarrow \infty$). Both of them are experimentally manageable. 

Under those conditions for a good quality QD at temperature below the nitrogen boiling point (77 K) \cite{Wei}, one should have $\Gamma \ll G = g \sqrt{\tilde{n}} < \Omega = g \sqrt{\tilde{m}} $, where $\Gamma \equiv \frac{1}{\tau}  $ stands for the total decay rate (inverse of the exciton lifetime $\tau$), and $2\Omega$ ($2 G$) is the Rabi frequency associated to the field A (B).

Hence, the matrix block corresponding to each $N$-plet resembles the representation of the position operator in the harmonic oscillator energy eigenstates \cite{Ramirez2013,QMbook}. Regarding the matrix blocks mixing $N$-plets, the larger the laser A respect to laser B, the more negligible such a coupling. 

Then, because $\hat{x} \mid x \rangle = x \mid x \rangle $, for each $j$-th $N$-plet  it is found that $\frac{1}{\sqrt{2}} \hbar g (\hat{b} \hat{\sigma}_+ + c.c.) \mid  j,N,\lambda_j  \rangle = \frac{\hbar g \lambda_j}{\sqrt{2}} \mid j,N,\lambda_j \rangle$  (with $j=+,-$ and $\lambda_j = j \lambda$).  The eigenenergies can thus be approximated according to $E_l \rightarrow E_{\lambda_j} \equiv  \hbar g \lambda_j$, where the continuous parameter $\lambda_j$ emulates the position eigenvalue of an dimensionless harmonic oscillator. In turn, the superposition coefficients can be approximated following  $c_{\tilde{n},j}^{N,l} \equiv \langle \phi_{\tilde{n},j}^N \mid \psi_l^N \rangle \rightarrow \langle   \phi_{\tilde{n},j}^N \mid j,N,\lambda_j \rangle \equiv \psi_{\tilde{n},j}^* (\frac{\lambda_j}{\sqrt{2}}) $, where 

\begin{equation}
\psi_{\tilde{n},j}(\frac{\lambda_j}{\sqrt{2}}) =  \left( \frac{1}{\sqrt{\pi} 2^{\tilde{n}} \tilde{n}!}\right)^{1/2} \exp \left(- \frac{1}{2} \frac{\lambda_j}{\sqrt{2}} \right) H_{\tilde{n}} (\frac{\lambda_j}{\sqrt{2}}) \hspace*{1ex} ,
\label{HOwavefunction}
\end{equation}

i.e. the wave function associated to the $\tilde{n}$-th energy eigenstate of the quantum harmonic oscillator ($H_{\tilde{n}} (x)$ is the $\tilde{n}$-th Hermite polynomial) \cite{Freedhoff,Ramirez2013}. 

Evidently, these eigenstates of the double dressed system satisfy orthonormality and completeness relations, which are used in finding the non-vanishing
matrix elements of the decaying transition operator between consecutive JCL rungs. They yield 

\begin{subequations}
	\begin{equation}
	\langle \pm , N-1 , \lambda_{\pm} \mid \hat{\sigma}_- \mid \pm , N , \lambda'_{\pm} \rangle = \pm \frac{1}{2} \delta (\lambda - \lambda') \hspace*{1ex} ,
	\label{transition1}
	\end{equation}
	\begin{equation}
	\langle \pm , N-1 , \lambda_{\pm} \mid \hat{\sigma}_- \mid \mp , N , \lambda'_{\mp} \rangle = \mp \frac{1}{2} \delta (\lambda + \lambda') \hspace*{1ex} ,
	\label{transition2}
	\end{equation}
\end{subequations}

where the sign difference in the delta functions has strong implications in the emission spectrum. Taking as reference the center of the energy miniband corresponding to each $N$-plet, equations (\ref{transition1}) and (\ref{transition2}) imply that transitions between minibands characterized by equal $j$, produce much sharper emissions than those between minibands with different $j$. Figure 2 illustrates this effect. There, green lines represent contributions to the central peak, while red (blue) lines contribute to a lower (higher) sideband, in the resonance fluorescence spectrum as measured by Y. He et al. in reference \cite{He2015}.


In this limit of very high $N$, by taking the real part of the Fourier transform of the correlation function for the dipole-moment operator \cite{emissiontheory,Freedhoff}, the resonance fluorescence spectrum of a monochromatically doubly dressed two-level system is given by


\begin{eqnarray} 
L_d(\omega)&=&\frac{\Gamma}{4 \pi}\Big\{\frac{\Gamma/2}{(\omega-\omega_{L})^2+(\Gamma/2)^2} + \frac{1}{4}\int_{-\infty}^{+\infty} d\lambda|\psi_{\tilde{n}}\left(\frac{\lambda}{\sqrt{2}} \right)|^2  
\Big[ \frac{3\Gamma/4}{(\omega-\omega_{L}-2\Omega- \lambda g)^2+(3\Gamma/4)^2} \nonumber \\
&& \hspace{5ex}	  +\frac{(3\Gamma/4)}{(\omega-\omega_{L}+2\Omega - \lambda g)^2 +(3\Gamma/4)^2} \Big] \Big\} \hspace*{1ex} ,
\label{spectrum}
\end{eqnarray}


for some $\tilde{n}$ in the weak laser.

\begin{figure}[h]
	\centering
	\includegraphics[width=10.5cm]{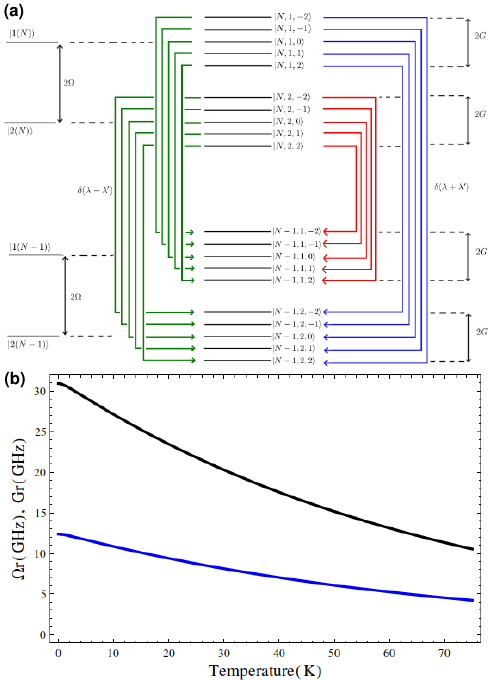}
	\caption{(a) Allowed dipole transitions between $N$-plets for consecutive JCL rungs. (b) Temperature dependence of renormalized Rabi frequencies $\Omega_r$ (black line)
		and $G_r$ (blue line).}
	\label{schematictransitions}
\end{figure}

Such spectrum has three components: A central peak, with a FWHM of $\Gamma$; and a lower (higher) energy side band, formed by the convolution of $\tilde{n}$ peaks with  FWHM of $\frac{3 \Gamma}{2}$, associated to the red (blue) transitions in figure \ref{schematictransitions}(a). 


\section{Thermal effects}

It has been experimentally observed by Y. Wei et al. in reference \cite{Wei}, that thermal effects on the emission spectrum of a driven quantum dot are entirely captured by a renormalization of the zero-temperature Rabi frequency and the accounting of a temperature-induced damping rate to be added to the radiative rate $\Gamma$.

Then, to evaluate the phonon effects on the emission spectrum of a doubly driven QD, we first consider separately the effects of the lattice vibrations on the emitter under the excitation of either laser. This is done by means of the master equation of a singly driven QD, which allows us to obtain the corresponding phonon-induced decoherence rate and Rabi frequency adjustment.

By applying the Born-Markov approximation \cite{footnote1,mahan1990many,wurger1998strong}, the master equation that governs the dynamics of the system within the polaron framework under resonant excitation \cite{mccutcheon2010quantum}, is given by 

\begin{eqnarray}
\frac{d\hat{\rho}(t)}{dt}=-\frac{\mathrm{i}}{2}[\tilde{\Omega}_r\hat{\sigma}_x,\hat{\rho}(t)]-
\frac{\tilde{\Omega}^2}{4}\int_0^{\infty} \mathrm{d}\tau\bigg([\hat{\sigma}_x,\hat{\sigma}_x\hat{\rho}(t)]
\Lambda_x(\tau) +\cos (\tilde{\Omega}_r \tau)[\hat{\sigma}_y,\hat{\sigma}_y\hat{\rho}(t)]\Lambda_y(\tau) \nonumber \\
+\sin(\tilde{\Omega}_r \tau)[\hat{\sigma}_y,\hat{\sigma}_z\hat{\rho}(t)]\Lambda_y(\tau) + h.c. \bigg),
\label{505}
\end{eqnarray}

where $\hat{\sigma}_x=(|X\rangle \langle0|+|0\rangle \langle X|)$,  
$\hat{\sigma}_y=\mathrm{i}(|0\rangle \langle X|-|X\rangle \langle0|)$, and
$\hat{\sigma}_z=(|X\rangle \langle X|+|0\rangle \langle0|)$, are the Pauli operators. $\tilde{\Omega}$ is half the Rabi frequency associated to the respective exciting laser and $\tilde{\Omega}_r=\tilde{\Omega} B$ is such a frequency renormalized by the expected value of the lattice displacement operator 

\begin{equation}
B \equiv \mathrm{exp}\bigg[{-\frac{1}{2}\int_0^{\infty}\mathrm{d}\omega \frac{J(\omega)}{\omega^2}
	\coth(\beta \omega /2)}\bigg],
\label{5.5}
\end{equation}

with $\beta=1/(k_B T)$, and $J(\omega)$ the phonon spectral density.  

Because the electron moves through the crystal lattice of the QD and its vicinity, this generates a deformation in the nuclei array, resulting in a polaronic system. In the continuum limit \cite{NazirJoPCM}, the system-bath interaction is characterized by the spectral density 
function

\begin{equation}
J(\omega) \equiv \frac{V}{\left( 2 \pi \right)^3} \int d\bm{k} \eta_{\bm{k}}^2 \delta\left( \omega - \omega_{\bm{k}} \right) = \eta \omega^3 \mathrm{e}^
{(-\frac{\omega}{\omega_c})^2},
\label{504}
\end{equation}

that describes the coupling of carriers to longitudinal acoustic (LA) phonons via deformation potential, in terms of the cutoff frequency $\omega_c$ \cite{krummheuer2002theory,RoyPRX} and of the constant $\eta$, which summarizes the strength of the exciton-phonon interaction and depends on the QD characteristics.

In turn, the correlations of the bath displacement operator are expressed as

\begin{equation}
\Lambda_x(\tau)=\frac{B^2}{2}(\mathrm{e}^{\phi(\tau)}+\mathrm{e}^{-\phi(\tau)}-2),
\label{5.1}
\end{equation}
\begin{equation}
\Lambda_y(\tau)=\frac{B^2}{2}(\mathrm{e}^{\phi(\tau)}-\mathrm{e}^{-\phi(\tau)}),
\label{5.2}
\end{equation}

where the phonon correlation function $\phi(\tau)$ \cite{roy2012polaron}, is defined according to 

\begin{equation}
\phi(\tau)=\int_0^{\infty}\mathrm{d}\omega \frac{J(\omega)}{\omega}(\cos (\omega \tau)
\coth(\beta \omega/2)-\mathrm{i}\sin(\omega \tau)).
\label{5.1-2}
\end{equation}








\subsection{Damping rates}

Following the treatment by McCutcheon and Nazir in reference \cite{mccutcheon2010quantum}, the damping rates in the polaron and weak coupling frameworks are obtained from the master equation (\ref{505}). Thus, the relevant fundamental rates that determine the evolution of the Bloch vector are

\begin{equation}
\Gamma_y=\frac{\tilde{\Omega}^2}{2}\gamma_x(0)
\label{506}
\end{equation}

and

\begin{equation}
\Gamma_z=\frac{\tilde{\Omega}^2}{4}(\gamma_y(\tilde{\Omega}_r)+\gamma_y(-\tilde{\Omega}_r)+2\gamma_x(0)),
\label{507}
\end{equation}

where 

\begin{equation}
\gamma_l (\omega)=2\text{Re}[K_l(\omega)],
\label{508}
\end{equation}

for ($l=x,y$), is given in terms of the polaron response function 

\begin{equation}
K_l(\omega)=\int_0^{\infty}\mathrm{d}\tau \mathrm{e}^{\mathrm{i}\omega\tau}\Lambda_l
(\tau).
\label{509}
\end{equation}

\subsubsection{Weak coupling rate}

Applying the weak coupling limit where $B \rightarrow 1$ and $\eta \omega_c^2 \ll 1$ (and consequently $\mathrm{e}^{\pm\phi(\tau)} \approx 1 \pm \phi(\tau)$, $\Lambda_x (\tau) \rightarrow 0 $ and $\Gamma_y \rightarrow 0 $) \cite{mccutcheon2010quantum}, the corresponding correlation function takes the form 


\begin{equation}
\Lambda_y (\tau) \rightarrow \Lambda_W(\tau)=\int_0^{\infty}\mathrm{d}\omega J(\omega)(\cos (\omega \tau) \coth(\beta \omega/2)-\mathrm{i}\sin(\omega \tau)),
\end{equation}

and the weak-coupling damping rate becomes 

\begin{equation}
\Gamma_z \rightarrow \Gamma_W=\frac{\pi}{2}J(\tilde{\Omega})\coth(\beta \tilde{\Omega}/2).
\label{511}
\end{equation}

This rate, in the high-temperature regime, varies linearly with the temperature  and does not take into account any renormalization of the Rabi frequency \cite{PRL-Rabi-2010-1,PRL-Rabi-2010-2}. 

\subsubsection{Polaron rate}

In the polaron theory, the total damping rate is given by the addition of the decay and decoherence rates $\Gamma_p=\Gamma_y+\Gamma_z$ \cite{mccutcheon2010quantum}, so that from equations (\ref{506}) and (\ref{507}), we have



\begin{equation}
\Gamma_p=\frac{\tilde{\Omega}^2}{4}\left[\gamma_y(\tilde{\Omega}_r)
+\gamma_y(-\tilde{\Omega}_r)+4\gamma_x(0)\right].
\label{e01fv}
\end{equation}

Inserting equations (\ref{508}) and (\ref{509}) into equation (\ref{e01fv}), the polaron damping rate turns in 

\begin{equation}
\Gamma_p=\frac{\tilde{\Omega}^2}{4} \Bigg\{ 2\text{Re} \left[ \int_0^{\infty}\mathrm{d}\tau
\mathrm{e}^{\mathrm{i}\tilde{\Omega}_r\tau}\Lambda_y(\tau) \right] + 2\text{Re} \left[\int_0^{\infty}\mathrm{d}\tau
\mathrm{e}^{-\mathrm{i}\tilde{\Omega}_r\tau}\Lambda_y(\tau)\right]+4\left[2\text{Re}\left(\int_0^{\infty}
\mathrm{d}\tau\Lambda_x(\tau)\right)\right] \Bigg\} ,
\end{equation}

which explicitly written in terms of the phonon correlation function, according to equations (\ref{5.1}) - (\ref{5.1-2}), is

\begin{eqnarray}
\Gamma_p &=& \frac{\tilde{\Omega}_r^2}{4}\bigg(\text{Re}\bigg[\int_0^{\infty}\mathrm{d}\tau
\mathrm{e}^{\mathrm{i}\tilde{\Omega}_r\tau}(\mathrm{e}^{\phi(\tau)}-\mathrm{e}^{-\phi(\tau)})
\bigg]
+\text{Re}\bigg[\int_0^{\infty}\mathrm{d}\tau
\mathrm{e}^{-\mathrm{i}\tilde{\Omega}_r\tau}(\mathrm{e}^{\phi(\tau)}-\mathrm{e}^{-\phi(\tau)})
\bigg] \nonumber \\
&& \hspace{7ex} +4\text{Re}\bigg[\int_0^{\infty}
\mathrm{d}\tau(\mathrm{e}^{\phi(\tau)}+\mathrm{e}^{-\phi(\tau)}-2)\bigg]\bigg).
\end{eqnarray}

\subsection{Single-phonon polaron rate}

There is an intermediate level of procedure between the weak-coupling and the full polaron approaches \cite{Wei}, so-called single-phonon approximation, which consists in taking for $\mathrm{e}^{\pm \phi(\tau)}$ the same truncated expansion as in the weak-coupling limit, but assuming that $B$ is still significant to make noticeable the renormalization of the Rabi frequency. Under such assumptions, the damping rate becomes 


\begin{equation}
\Gamma_{1-ph}=\frac{\pi}{2}J(\tilde{\Omega}_r)\coth(\beta \tilde{\Omega}_r/2).
\label{5.3}
\end{equation}

The $\Gamma_W$, $\Gamma_{ph-1}$ and $\Gamma_p$ rates are expected to have similar 
behavior at low temperatures ($T<15$ K), and then differ noticeably as the bath-induced fluctuations and multiphonon effects become important at higher temperatures.


\section{Results}

\begin{figure}[h]
	\centering
	\includegraphics[width=10.5cm]{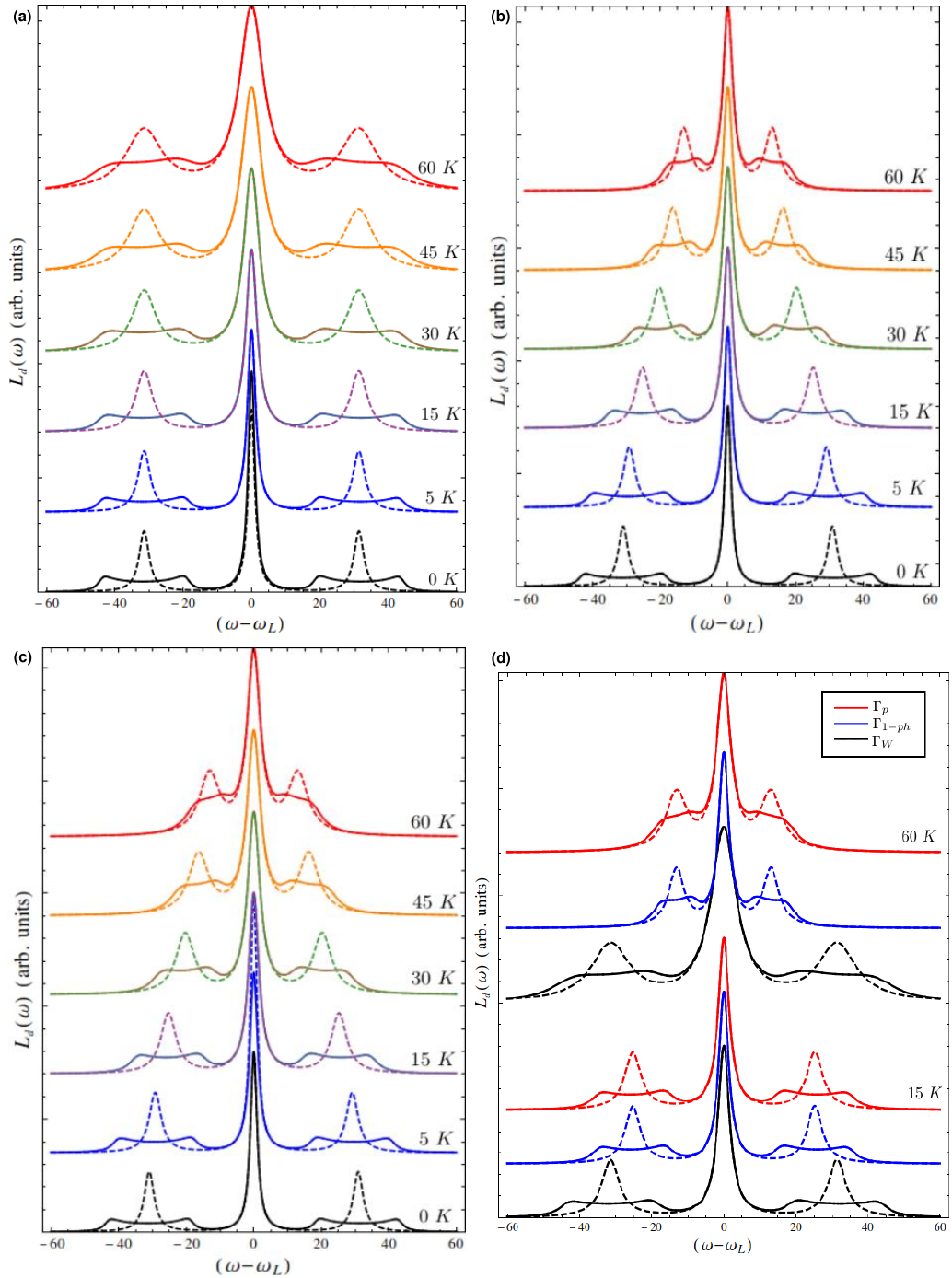}
	\caption{Temperature dependent phonon effects on the resonance fluorescence of a QD driven by one (dashed lines) and two (solid lines) electromagnetic fields in the (a) weak-coupling limit $\Gamma_W$, (b) the single-phonon limit, and (c) the full polaron framework. (d) Comparison of the phonon effects on the photon emission from a QD driven by one (dashed lines) and two (solid lines) electromagnetic fields at $15$ K and $60$ K, as simulated within the different approximation levels here studied: weak-coupling (black line), single-phonon (blue line) and polaron (red line) approaches.
	}
	\label{5-31}
\end{figure}

For the doubly driven system under study, we now include the phonon effects on the resonance fluoresce spectrum by considering the modification of the decay rate and Rabi frequencies appearing in equation (\ref{spectrum}). Hence, on the one hand the radiative broadening $\Gamma$ is substituted by the total decay rate $\Gamma_T = \Gamma + \Gamma_i$ for the cases $i=W,p,1-ph$, while on the other hand the original Rabi frequencies $2 \Omega$ and $2 G$ are substituted by the renormalized ones $2 \Omega_r \equiv 2 B \Omega$ and $2 G_r \equiv 2 B G$. 

In the numerical simulations, we use parameters consistent with those reported in reference \cite{Wei}, corresponding to experiments on single self-assembled InAs/GaAs QDs embedded in a microcavity  with  low $Q$ factor ($\sim 200$). Thus, the Rabi frequencies are tuned to $2\Omega = 10 \pi$ GHz and $2G = 4 \pi$ GHz ($\alpha^4 \approx 0.025$), respectively. The spontaneous emission rate of the neutral exciton state is taken as $\Gamma=2.35$ GHz, the coupling exciton-phonon constant as $\eta=2.535$ x $10^{-7}$ GHz$^{-2}$ and the phonon cutoff frequency as $\omega_c = 493.33$ GHz.

Figure 2(b) shows the temperature-dependent renormalization of the original Rabi frequencies $2\Omega$ and $2G$, correspondingly. There can be observed how in the range $0-70$ K, the renormalized Rabi frequencies are almost reduced by a factor $3$. It is worth noting that while in the weak-coupling approach this effect is despised, its amount is the same in the polaron and single-phonon approaches.   

The doubly driven QD emission spectra, simulated within the three different levels of approximation considered in this work, are shown in figures 3(a) - 3(c) for a temperature range between $0$ and $60\ K$ \cite{comment-2}. The singly driven case is also shown in dashed lines for comparison.  

At zero temperature, where the thermal dissipation is absent, it can be observed how turning on the second driving laser modifies the side-peaks into optically active side-bands. The width of those bands corresponds to the Rabi frequency associated to the weaker laser and then is controllable by adjusting its power. As long as this Rabi frequency is significantly smaller than the one associated to the stronger laser (which corresponds to the distance between the center of the side-bands and the central peak), there will be distinguishable optically dark frequency ranges between the bright side-bands and the central peak. The greater the ratio between the Rabi frequencies, the wider those dark ranges.  Such tunability is precisely what makes the system very promising for optical switching \cite{Ramirez2013,He2015}.

As the temperature increases, the sharpness of the side-band limits is blurred while the central peak broadens, 
eventually extinguishing the dark regions between them. According to the three studied models, for temperatures above 45 K the side-bands and central peak are so overlapped that the alternating bright-dark regions are practically undone by the thermal dissipation. However, at temperatures up to 30 K, the contrast between optically active and inactive frequencies along the emitting window is greater than $75\%$, as compared to the 0 K case. Thus, from the simulated spectra the plausibility of using a doubly driven emitter as an efficient optical frequency selector at the temperature range 0 - 30 K, can be anticipated.    

Figure 3(d) compares the emission spectra obtained from the weak coupling, the single-phonon and the full-polaron models, at 15 K and at 60 K. There, overestimation of the thermal effects calculated under the weak coupling model is evidenced \cite{Oscar2018}. Meanwhile, the very similar features observed in the spectra simulated within the single-phonon and the polaron models reveal that the former is accurate enough for describing the thermal influence in the studied temperature scope. In contrast with the results from the weak-coupling model, a substantial reduction of the overall optically active frequency window is appreciated in the spectra obtained from the single-phonon and polaron models in which the Rabi frequency renormalization is included.

\section{Summary and conclusions}

In this work, the thermal dissipation effects on the resonance fluorescence of an artificial atom simultaneously driven by two incoherent electromagnetic fields were computationally studied. By expanding a previously existing model that describes a doubly driven atom perfectly isolated from phonons, we included the  temperature dependence unavoidably present in solid-state based quantum emitters. Thus we developed the necessary framework to simulate the temperature dependence of the photon emission from this highly tunable system, that allows observation and manipulation of quantum interference phenomena in mesoscopic systems.    

Within three different levels of approximation for the carrier-phonon coupling and using realistic material parameters, we simulated the resonance fluorescence spectra of a doubly driven artificial atom embedded in a acoustic phonon environment in the temperature range $0-60$ K. As a main result, our calculations show robustness to phonon dissipation of the efficient control of the density of optical states up to neon boiling temperatures ($\sim30$ K), making the scheme promising for applications in cryogenic optoelectronic devices requiring frequency filtering.     


\section{Acknowledgments}
The authors acknowledge the Research Division of UPTC for financial support through project No. SGI-2527. H.Y.R. also acknowledges support from the Fulbright program for Colombian Scholar Visitors in USA. 





\bibliographystyle{unsrtnat}

\bibliography{final-bibliography-thermal-effects-double-driving}

\begin{thebibliography}{56}
\bibitem[foo()]{footnote1}
The use of this approximation is appropriate here because along the studied
  temperature range, the thermal energy is much smaller than the characteristic
  transition energy of the emitter and the excitation is strong enough to make
  thermalization much faster than the relevant optical dynamics.
\bibitem[com()]{comment-2}
Because the laser b is significantly weaker than laser a, its contribution to
  the damping rates is neglected and only the renormalization of the
  corresponding rabi frequency is considered.
\end{thebibliography}

\end{document}